\documentclass[%
 reprint,
footinbib,
amsmath,amssymb, 
aps,
prl
]{revtex4-2}

\usepackage{graphicx}
\usepackage{dcolumn}
\usepackage{bm}
\usepackage{hyperref}
\usepackage[mathlines]{lineno}
\usepackage{mathtools}
\usepackage{wrapfig}
\usepackage{graphicx} 
\usepackage{setspace}

\usepackage[utf8]{inputenc}
\usepackage{lipsum} 
\usepackage{xcolor}

\usepackage{bm}

\begin{document}


\title{Genetic interfaces at the frontier of expanding microbial colonies}

\author{Jonathan Bauermann}
\author{David R. Nelson}%
\affiliation{%
 Department of Physics, Harvard University, Cambridge, MA 02138, USA
}
\begin{abstract}
We study the genetic interfaces between two species of an expanding colony that consists of individual microorganisms that reproduce and undergo diffusion, both at the frontier and in the interior.
Within the bulk of the colony, the genetic interface is controlled in a simple way via interspecies interactions. However, at the frontier of the colony, the genetic interface width saturates at finite values for long times, both for neutral strains and interspecies interactions such as antagonism. 
This finite width arises from geometric effects: genetic interfaces drift toward local minima at an undulating colony frontier, where a focusing mechanism induced by curvature impedes diffusive mixing.
Numerical simulations support a logarithmic dependence of the genetic interface width on the strength of the number fluctuations.
\end{abstract}

\date{\today}

\maketitle

Via an interplay of diffusion and reproduction, biological colonies can invade new areas and expand. This dynamics, which often takes place in two dimensions, includes saturation for high densities in the colony interior, and can be described by the celebrated Fisher–Kolmogorov–Petrovskii–Piskunov (FKPP) equation~\cite{Fisher1937, Kolmogorov1937}, which was introduced nearly a century ago. Since then, the FKPP equation has become a foundational tool in mathematical biology and spatial population dynamics, capturing the essence of population range expansion in the form of pulled traveling-wave solutions~\cite{Murray2002}.
Despite the success of this minimal model, many aspects of such "proliferating active matter" remain poorly understood~\cite{Hallatschek2023}. 
For example, it has been shown that mechanical forces between nearby cells can alter colony growth~\cite{Farrell2013,Giometto2018}. Another example is provided by bacterial colonies expanding in three dimensions, where "growth-induced" instabilities can lead to "broccoli-like morphologies"~\cite{MartnezCalvo2022}. 
Another initially surprising result, at least from the perspective of phase separation of binary mixtures, is the observation that when a well-mixed population of two neutral non-motile microbial species - for example, two identical microbial strains with two different heritable genetic labels - is placed on a Petri dish, the colony demixes as it expands~\cite{Hallatschek2007}. 
However, understanding the expansion of these colonies with non-motile microbes relies on reproduction primarily at the frontier. As individuals are born, they are described by an Eden-like growth model~\cite{Saito1995} and the stability of genetic interfaces at the frontier, even neutral ones, becomes more understandable.

In this work, we study with agent-based simulations the width of the genetic interface between two species at the frontier of an expanding colony composed of motile cells that can mix dynamically via diffusion everywhere, both in the interior and at frontiers. This problem is more subtle:
We show that this width is controlled by a combination of the drift of the interface to local minima of an undulating frontier, where the diffusive mixing is overcome by a geometric focusing effect.
We model here the expansion of such colonies via the FKPP equation with neutral, antagonistic, and mutualistic interspecies interactions, eventually including the effect of demographic noise.
While the dynamics in the bulk depends strongly on these interspecies interactions, surprisingly, the genetic interface at the colony's frontier is largely independent of the nature of these interactions. 
Radially expanding colonies with two neutral genotypes and diffusion everywhere were previously modeled for radial range expansion in Ref.~\cite{Hallatschek2010}. 
However, the connection between the interface widths, undulations and genetic drift was not studied.
To keep analysis simple, we avoid the complications of inflation by considering linear initial conditions, along the lines of the "razor blade" innoculations of Ref.~\cite{Hallatschek2007}.

\textit{Deterministic dynamics - } We begin by neglecting genetic drift, i.e., demographic noise, and studying the deterministic dynamics for flat fronts. The dynamics of the concentration fields $c_i$ of two species $i=A,B$ evolve according to two coupled FKPP equations,
\begin{equation}
\label{eq:FKPP2}
    \partial_t c_i = D \nabla^2 c_i  + \mu c_i  (1  - c_A - c_B + \epsilon  c_j) \, ,
\end{equation}
with $j\neq i$ \cite{Pigolotti2013}. For simplicity, we have chosen an identical diffusivity $D$ and overall reproduction rate $\mu$ for both species, leading to nearly identical growth dynamics at the frontier of any colony where the total concentration $c_T = c_A + c_B \ll 1$. 
However, under crowded conditions, i.e., $c_T \approx 1$, we allow for symmetric cross-species interactions, parameterized by $\epsilon$, which is assumed to be small, $|\epsilon| \ll 1$.
For $\epsilon=0$, the two species are genetically neutral even under crowded conditions, while for $\epsilon \neq 0$, the two species can interact mutualistically ($\epsilon>0$) or antagonistically ($\epsilon<0)$.

\begin{figure}[t!] 
    \includegraphics[width=0.8\linewidth]{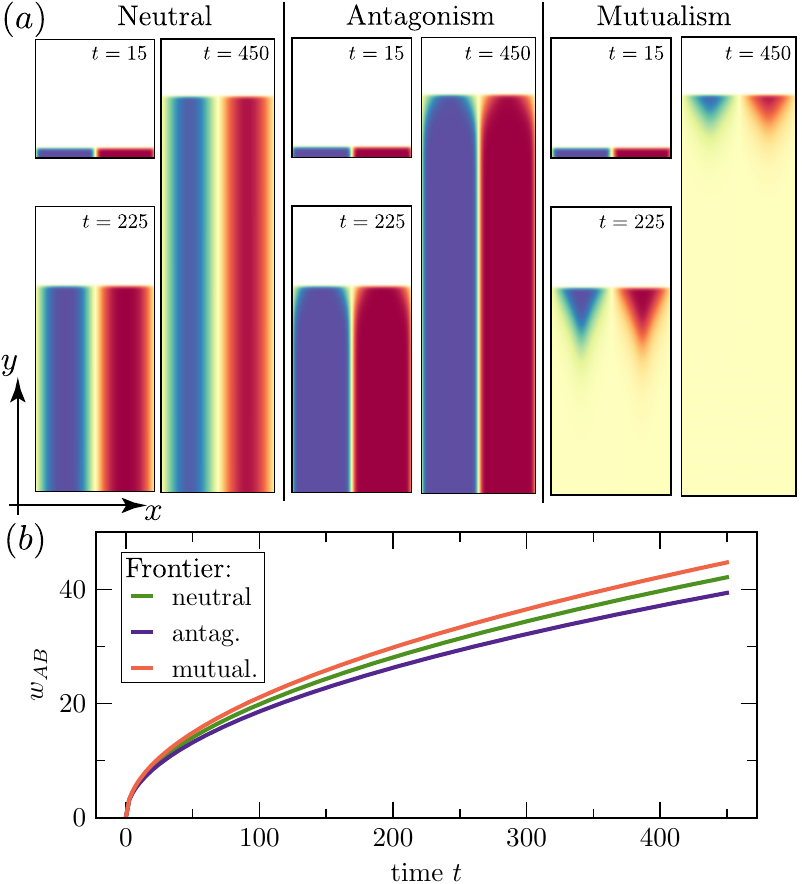}
    \caption{
    \textbf{Deterministic FKPP dynamics for two interacting species}:\label{fig:FKPP} 
    (a) Concentration profiles at three different times $t$ for neutral/antagonistic/mutualistic interactions, with $\epsilon =0,-0.1,0.1$. Periodic boundary conditions are employed in the $x$-direction.
    (b) $AB$-interface width $w_{AB}$ along the x-direction at the frontier of the colonies (computed along the value of $y$ such that $c_T=0.5$) for these three interaction values. Units: $\lambda = \sqrt{D/\mu}$ (length) and $\tau=1/\mu$ (time);
    System-dimension: $L_X=256$, $LY = 1024$; 
    Grid-resolution: $512 \times 2048$ points in the $(x,y)$-plane; with numerical time-step $dt=0.005$.
    } 
\end{figure}

For neutral interactions ($\epsilon=0$), we can analytically solve for the expansion of the colony when both species are initialized next to each other with a flat frontier. 
We assume $c_A(x,y,0) = \Theta(y_0 - y) \Theta( x_0 - x)$ and $c_B(x,y,0) = \Theta(y_0 - y) \Theta( x-x_0)$ at $t=0$, where $\Theta(z)$ is the step function. 
The dynamics of total concentration $c_T$ then obeys the classical one-dimensional FKPP equation $
\partial_t c_T = D \partial_y^2 c_T + \mu c_T (1-c_T) $ in the $y$ direction. Thus, a pulled traveling wave with $c_T(x,y,t)= c_T(y-vt)$ and a finite interface width is established, where $c_T(z)$ is a traveling wave solution. 
No specific traveling wave solution and its corresponding velocity are uniquely fixed by Eq.~\eqref{eq:FKPP2}. However, for the step-like initial condition in the $y$ direction, it is known that the velocity reaches $v=v_\text{FKPP} = 2\sqrt{D \mu}$ after an initial relaxation period \cite{Murray2002}.
Upon invoking a separation ansatz, we find
 \begin{equation}
    c_i(x,y,t) =  \frac{c_T(y-vt)}{2} \left[ 1 \pm \text{erf}\left( -\frac{x-x_0}
    {2 \sqrt{D t}} \right) \right] \; ,\\
    \label{eq:sol_neut}
\end{equation}
for $i=A,B$, where $\text{erf}(z)$ is the error-function, entering with a $+$ sign for $i=A$ and a $-$ sign for $i=B$.
Thus, as the interface advances in the $y$-direction, the initial sharp genetic interface between $A$ and $B$ in the $x$-directions broadens diffusively over time, with a width of $w_{AB}\propto \sqrt{t}$. This broadening is identical both in the bulk and at the frontier of the colony.

Numerical results for the neutral setting described above and also for the antagonistic and mutualistic mixtures with identical initial conditions are shown in Fig.~\ref{fig:FKPP}(a). 
With the flat front initial conditions described above, where $c_A(x,y=0,t=0)=\Theta(x-x_0)$, $c_B(x,y=0,t=0)=\Theta(x_0-x)$, the colony is allowed to expand upwards.
For the neutral setting $\epsilon=0$, corresponding to our analytical solution, a diffusive mixing along $x$ independent of the $y$-position within the colony appears and is well-described by Eq.~\eqref{eq:sol_neut}.
However, for mutualistic or antagonistic interspecies interactions, the bulk dynamics change dramatically.
The colony of an antagonistic mixture ($\epsilon<0$) establishes a finite interface along $x$ with width $w_{AB} = 2 \sqrt{D/(\mu |\epsilon|)}$ deep down in the colony, while in a mutualistic mixture ($\epsilon>0$), the well-mixed state with $c_A=c_B=0.5$ is a fixed point of the dynamics. This well-mixed state is established via a second pulled wave with velocity $v_\text{mut}=\sqrt{2 D \mu \epsilon}$, leading to a transient wedge of the pure states behind the frontier. For further details, see Appendix~A. Note that $v_\text{mut}<v_\text{FKPP}$ when $\epsilon$ is small, as assumed here.

At the frontier, however, even for the antagonistic and mutualistic settings, the genetic interface between the species broadens over time; see Fig.~\ref{fig:FKPP}(b). 
Here, we measured the interface width at the frontier by first defining the frontier of the colony as the line where $c_T(x,y)= 1/2$. 
Along this line, we fit the function $\Delta(x) = (\text{erf}[(x-x_0)/w_{AB}])/2$ to the difference $\Delta = (c_A - c_B)/2$ around each interface.
This diffusive broadening increases like $\sqrt{t}$, and has only a weak dependence on the interspecies interaction strength $\epsilon$. 
Interspecies interactions are quite important for the bulk dynamics, while the frontier dynamics of the pulled wave, described by the FKPP equation, is controlled by the leading edge of the colony \cite{VANSAARLOOS2003}.

\textit{Stochastic dynamics - } 
To introduce number fluctuations, we use a birth–death process similar to \cite{Doering2003, Pigolotti2013} on a triangular lattice with one of the three principal nearest neighbor directions aligned with the $x$-axis.
Every individual $X_i$ can undergo a
\begin{align}
    &\text{birth: } X_i \rightarrow 2 X_i, \text{ rate } \mu , \text{ or }\label{eq:birth}\\
    &\text{death by competition: } X_i + X_j \rightarrow X_j, \text{ rate } \lambda_{ij} , \label{eq:death}
\end{align}
process, where $i,j = A,B$. 
The latter models the competition between individuals $X_i$ and $X_j$ on each lattice site.
Again, we focus for simplicity only on the symmetric case $\lambda_{ij}=\lambda_\text{self}$, when $i=j$ and $\lambda_{ij}=\lambda_\text{cross}$ when $i\neq j$.
In addition to the local birth–death process, every individual can jump with a rate $\rho$ to one of the six neighboring sites.
We used a Gillespie algorithm to simulate these processes~\cite{gillespie1977exact}.
The competition leads to a saturation of the deme-size $N_T = N_A + N_B$, where $N_i$ are the number of individuals of species $i=A,B$, on each lattice site.
If only one species is present at a lattice site, this saturation level, known as carrying capacity, is given by $\bar{N}_T = \mu/\lambda_\text{self}$.
Only for the neutral case of $\lambda_\text{cross}=\lambda_\text{self}$ is the carrying capacity independent of the composition of the population.
In the antagonistic case of $\lambda_\text{cross}>\lambda_\text{self}$, the carrying capacity is reduced, while for the mutualistic setting of $\lambda_\text{cross}<\lambda_\text{self}$, it is increased~\cite{Pigolotti2013}.
In the limit $\bar{N}_T\rightarrow \infty$, the variables $c_i = N_i/\bar{N}_T$ follow the deterministic dynamics stated in Eq.~\eqref{eq:FKPP2} with $\epsilon = 1- \lambda_\text{cross}/\lambda_\text{self}$, and $D=3\rho/8$; see Refs.~\cite{Doering2003, Pigolotti2013}.

\begin{figure}[!t] 
    \includegraphics[width=0.8\linewidth]{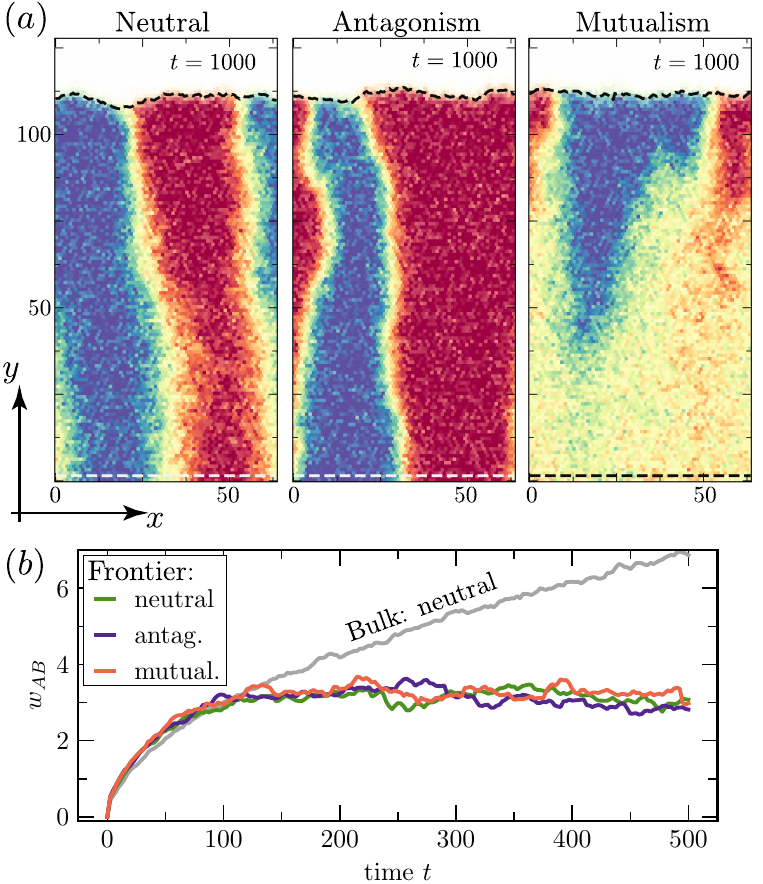}
    \caption{
    \textbf{Stochastic FKPP-dynamics for two interacting species, with and without interactions,}\label{fig:stoch} again with periodic boundary conditions in the $x$-direction
    (a) Lattice configurations of typical simulations for neutral/antagonistic/mutualistic interactions, with $\epsilon =0,-0.1,0.1$ with $t=1000$.
    (b) Genetic interface width $w_{AB}$ along the x-direction at the frontier of the colonies along a line where $N_T=\bar{N}_T/2$ (indicated by the black dashed lines in (a)) for these three interaction settings; we also show the width at the initialization height for the neutral setting as a function of time (indicated by the white or black dashed line at the bottom of (a)).
    After a brief transient, this quantity gives the bulk interface width deep in the colony interior. A typical deme size for these simulations is $\bar{N}_T=\mu/\lambda_\text{self}\approx 100$
    Parameters: $\mu = 0.1$, $\lambda_\text{self}=0.001$, $\lambda_\text{cross}= \lambda_\text{self}(1-\epsilon)$, $\rho=0.1$;  
    Lattice-dimension: $M_X=64$, $M_Y=128$; 
    } 
\end{figure}

In Fig.~\ref{fig:stoch}(a), we show lattice configurations for such simulations with $\bar{N}_T = 100$, for the same coarse-grained interactions as for the deterministic dynamics shown in Fig.~\ref{fig:FKPP}.
Again, we initialize segregated pure states with $N_{A/B} = \bar{N}_T$, at the bottom left/right (white or black dashed lines indicate the height of the initial colony). 
Within the bulk, we find a dynamics similar to the dynamics described by the deterministic problem ($\bar{N}_T \rightarrow \infty$) for the corresponding three cases: a diffusive spreading of the width $w_{AB} \propto \sqrt{t}$ for the neutral interactions (not shown); a finite interface width for the antagonistic interactions (not shown); and a well-mixed state, reached via a second, slower pushed wave behind the frontier for mutualistic interactions (also not shown).
However, at the frontier, defined as the line where $N_T= \bar{N}_T /2$ (indicated by the black dashed line), the dynamics of the genetic interface differs qualitatively from the deterministic case.
Again, by a fit to an error function profile to the difference $(N_A - N_B)/\bar{N}_T$ along the frontier (again subject to periodic boundary conditions in the $x$-direction, see Appendix~B for details), we find that, the sharp $AB$-interface initially broadens over time but eventually reaches a finite width, independent of the interactions, as shown in Fig.~\ref{fig:stoch}(b).
In addition to the frontier widths (in a comoving frame), we show the width $w_{AB}$ at the fixed height at the initialization for the neutral case (gray line). Even for times at which the width at the comoving frontier has reached its stationary value, the width at this height still grows $\propto \sqrt{t}$ in time. This broadening is representative of the bulk behavior after an initial transient, and similar to the deterministic case, shown in Fig.~\ref{fig:FKPP}.

\begin{figure}[!t] 
    \includegraphics[width=0.99\linewidth]{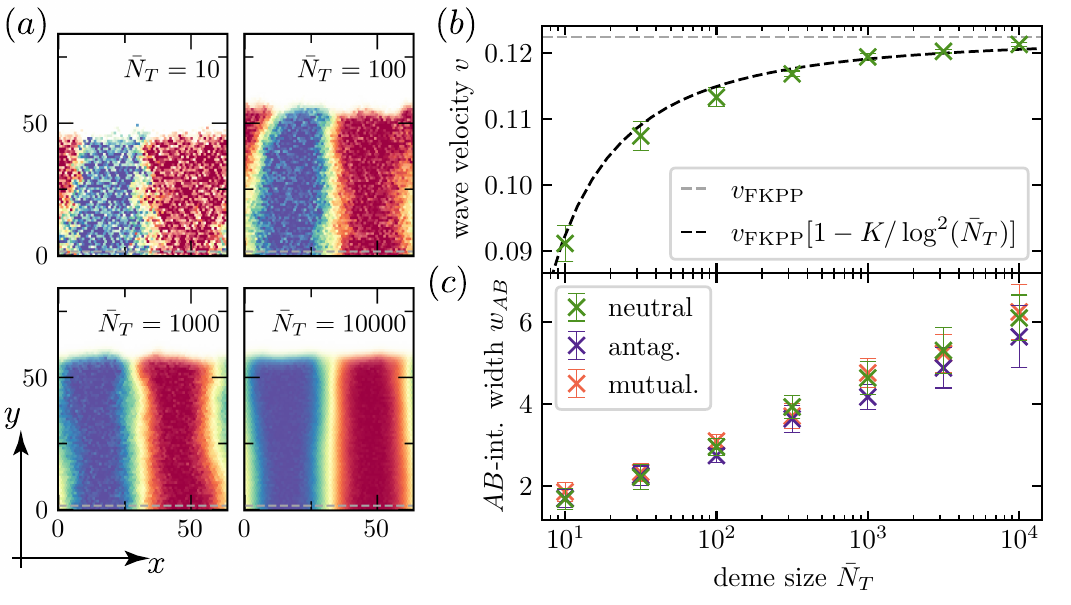}
    \caption{
    \textbf{Effects of noise strength}:
    (a) Lattice configurations of typical simulations for neutral interactions $\epsilon=0$ with $\bar{N}_T=10,100,1000,10000$.
    (b) Wave velocities for the pulled FKKP waves as a function of $\bar{N}_T$ with standard error (averaged over $20$ independent simulations) for neutral interactions, deterministic limit (gray dashed line), and correction due to fluctuations (black dashed line) as functions of the carrying capacity $\bar{N}_T$ with $K=1.30$ fit to the functional form $v_\text{FKKP} \sim v_\text{FKKP}(\infty)(1-K/\log^2(\bar{N}_T)$.
    (c) Interface width $w_{AB}$ along the frontier for interaction parameters $\epsilon=0,-0.1,0.1$ as functions of the carrying capacity $\bar{N}_T$, at time $t=500$.
    Parameters:
    $\mu = 0.1$, $\lambda_\text{self}=\mu/\bar{N}_T$, $\lambda_\text{cross}= \lambda_\text{self}(1-\epsilon)$, $\rho=0.1$;  
    Lattice-dimension: $N_X=64$, $N_Y=128$; 
    \label{fig:width_vs_N} 
    } 
\end{figure}
To further investigate the consequences of the number fluctuations embodied in genetic drift, we vary the carrying capacity $\bar{N}_T$, where we expect to recover deterministic results in the limit $\bar{N}_T \rightarrow \infty$. Figure~\ref{fig:width_vs_N}(a) shows examples of lattice configurations for neutral conditions with carrying capacities $\bar{N}_T = 10,100,1000,10000$.
All simulations were stopped after evolving the same step-like initial condition until $t=500$.
Note that the height of the colonies at this time depends on the deme size $\bar{N}_T$. This dependency results from a correction to the wave velocity $v\approx v_\text{FKPP}[1-K/\log(\bar{N}_T)]$, with $K>0$, as expected for the stochastic FKPP equation \cite{Brunet1997, Brunet2001, VANSAARLOOS2003, Panja2004}.
We test this expectation in Fig.~\ref{fig:width_vs_N}(b), where we show the numerically estimated values of $v$ and their standard error, both averaged over $ 20$ simulations for the neutral setting. 
In addition, we show the deterministic velocity $v_\text{FKPP}$ (gray dashed line) and the fluctuation-induced correction (black dashed line), mentioned above.

A novel finding of this work is that the genetic interface width $w_{AB} $ at the frontier depends logarithmically on the carrying capacity, i.e., $w_{AB} \propto \log(\bar{N}_T)$; see Fig.~\ref{fig:width_vs_N}(c). 
Furthermore, this dependency appears to be independent of the interspecies interactions. 
We also performed numerical simulations of the FKPP with simple demographic noise (a computationally simpler alternative to birth/death processes with descrete particles) and found the same logarithmic dependency of the genetic interface width; see Appendix~C.
We also find that due to number fluctuations when there is diffusion everywhere, each of the two genetic interfaces imposed by our periodic boundary conditions can occasionally split into three separate interfaces: schematically, $AAA|BBB \rightarrow AA|B|A|BB$.
We excluded the rare simulations where such splitting events occurred in the simulated time window of $t=500$; see Appendix~D for details.

\begin{figure}[!t] 
    \includegraphics[width=0.84\linewidth]{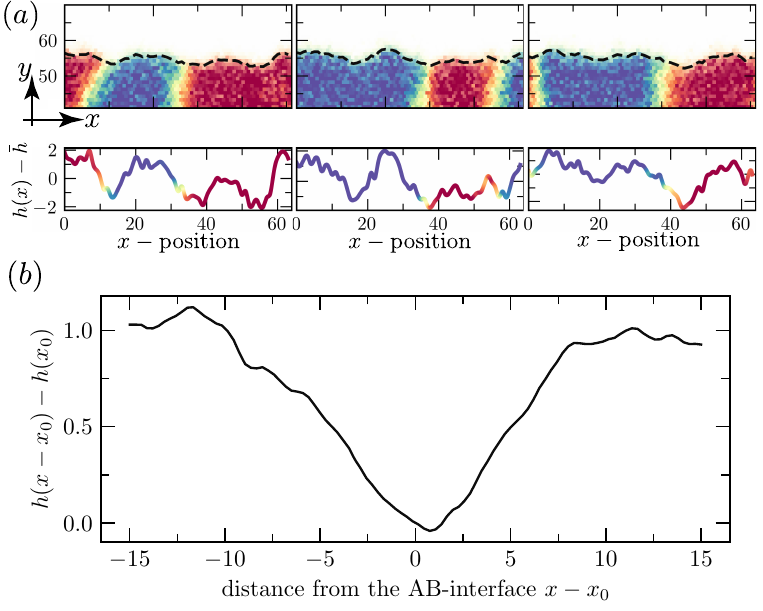}
    \caption{
    \textbf{$AB$-interfaces localize at height minima}:\label{fig:und_intpos} 
    (a) Lattice configurations at an undulating frontier for three typical simulations for neutral interactions with $\bar{N}_T=100$ (top row), with colony height indicated with a black dashed line, and colored with the local genotype fraction at time $t=500$ (bottom row), expanded vertical scale. Our periodic boundary conditions in the $x$-direction ensure that two genetic interfaces can be studied for each simulation.
    (b) Relative colony height around the $AB$-interface at position $x_0$, averaged over 20 individual runs and each of the two interfaces, given in units of the lattice spacings.
    Parameters: $\mu = 0.1$, $\lambda_\text{self}=\lambda_\text{cross}=0.001$, $\rho=0.1$;  
    Lattice-dimension: $N_X=64$, $N_Y=128$; 
    } 
\end{figure}

\textit{Undulations of the frontier - }
We now argue that the undulations of the colony frontier, a natural consequence of number fluctuations even if the frontier is initially flat, strongly influence the location of the finite interface widths between two different species we observe at the front. 
Indeed, we find that, on average, genetic interfaces tend to localize at minima of the frontier.
To demonstrate this, we show the interface region of three typical simulations with neutral interactions in Fig.~\ref{fig:und_intpos}(a). 
The lattice configurations are shown in the top row, while the height of the frontier in $y$-direction (indicated by the black dashed line on top of the lattice), colored in the local genotype fraction $f=N_A/(N_A+N_B)$, is shown in the bottom row, with an expanded vertical scale. 
In these three typical examples, the co-localization of the genetic interface with minima of the undulations can be seen. 
We quantify this tendency in Fig.~\ref{fig:und_intpos}(b), where we show the colony height profile $h(x) - h(x_0)$ relative to the height at the $AB$-interface position at $x_0$, averaged over both interfaces with 20 independent runs.

Thus, even for neutral interactions that do not discriminate between different species, the position of the genetic interfaces nevertheless typically lags behind the average position of the frontier. 
This remarkable focusing effect of frontier minima can be understood qualitatively by noting that any individual who happens to be slightly ahead of the average position of the colony has an advantage compared to the individuals on neighboring demes at the frontier, due to preferred access to uncolonized space.
Whether this individual is of species $A$ or $B$, a forward-bulged domain of this type emerges.
This downhill growth has already been described in the literature on the phase-ordering in colonies that expand stepping stone models without diffusion in the bulk~\cite{Drossel2000, Chu2019}. 
We argue here that this focusing effect at frontier minima is accompanied by a finite width $w_{AB}$ and impedes the diffusive mixing at the frontier in a way that depends strongly on the strength of number fluctuations.

\begin{figure}[!t] 
    \includegraphics[width=0.99\linewidth]{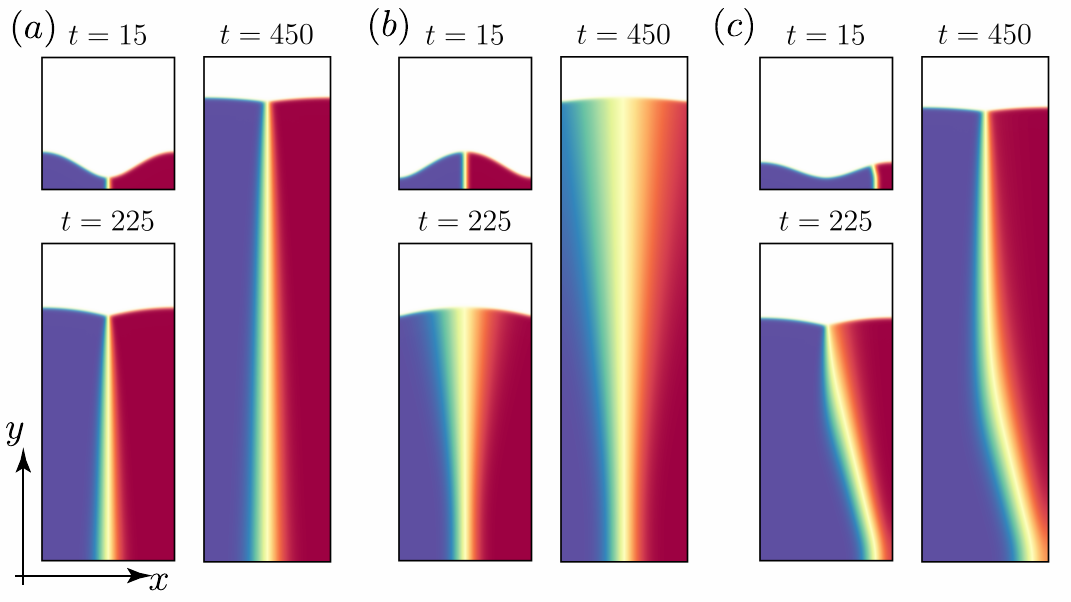}
    \caption{
    \textbf{Frontier undulations alter interface broadening}:\label{fig:undul} 
        Concentration profiles at three different times $t$ for neutral interactions $\epsilon =0$.
        Initially, the colony height has an imposed long wavelength $\cos(x/L_X)$ undulation. 
        In (a) and (b), the $AB$(i.e. blue/red)-interface is initially positioned as a step function at the minimum and maximum of the cosine function.
        Note the squeezing of the interface width at the frontier in (a), as opposed to the broadening of the frontier interface width in (b).
        (c) An $AB$-interface initially positioned off center from cosine indentation is attracted to the minimum of the slowly relaxing cosine undulation.
        Parameters same as in Fig.~\ref{fig:FKPP}, but no-flux boundary conditions in the $x$-, and $y$-direction, so that we can focus on a single interface.
    } 
\end{figure}
First, we demonstrate this focusing effect with deterministic dynamics (corresponding to the limit $\bar{N}_T \rightarrow \infty$, with no number fluctuations), where we perturb the initial colony height with a cosine undulation.
As this height perturbation slowly decays away, the focusing can be seen when the genetic interface remains at the minimum, see Fig.~\ref{fig:undul}(a). However, the interface at the frontier is much sharper than the usual diffusion broadening we find deep down in the bulk. 
In sharp contrast, when the genetic interface is positioned exactly at a maximum, it remains there but broadens faster than in the bulk, see Fig.~\ref{fig:undul}(b).
The tendency of the genetic interface to drift downhill towards frontier minima, with biased motion in the $x$-direction, is illustrated in Fig.~\ref{fig:undul}(c). Here, we initialized the genetic interface half way between the minimum and the maximum of the long-wavelength undulation embodied in the initial condition. 
First, the interface drifts downhill and broadens along the way. However, once it reaches the minimum, it gets squeezed at the minimum while the cosine perturbation of the frontier slowly decays away.

Importantly, when the frontier is forced to be flat with similar lattice models to those described above, the genetic interface at the frontier broadens in the same way as it does in the bulk; see Appendix~E.
Thus undulations, whether imposed artificially or the result of number fluctuations, are crucial for the squeezed interface widths we find in our simulations. 

\textit{Frontier model - }
When only interested in colony frontier dynamics, with negligible diffusion in the interior, phenomenological frontier models of the local genetic fraction $f(x,t)=c_A/(c_A + c_B)$ have proven to be useful~\cite{Korolev2010}. 
Here, instead of modeling the full two-dimensional dynamics at the frontier and the bulk, the fraction at the frontier is described by a one-dimensional equation that depends on time (a coordinate the locates the frontier position along the y-axis) and the coordinate $x$ along the frontier.
Recently, the interplay between a local selective advantage describing direct competition and the fitness advantage of faster reproduction under dilute conditions, leading to frontier deformations, has been studied. In these models, the fraction dynamics at the frontier $f(x,t)$ was coupled to the interface height $h(x,t)$ of the frontier ~\cite{ Horowitz2019, Swartz2023, Swartz2024}. 

As discussed earlier in our full two-dimensional simulations (see Fig.~\ref{fig:FKPP} and Fig.~\ref{fig:stoch}), the fraction dynamics $f(x,t)$ at the frontier of the colony following a pulled wave FKPP dynamics into unoccupied territory only depends weakly on interspecies interactions. 
Furthermore, in Eq.~\eqref{eq:FKPP2} and our modeling of number fluctuations, we assumed identical reproduction rates $\mu$ of both species in the dilute limit. 
We focus on the neutral model ($\epsilon=0$), where the dynamics of the height of the frontier, denoted as $h(x,t)$, is independent of the local fraction $f(x,t)$.
However, as we have argued above, the profile of the frontier clearly influences the dynamics of the fraction $f$.
In the lowest order of this coupling, an effective advecting drift appears in the fraction dynamics~\cite{Chu2019,Horowitz2019}:
\begin{gather}
    \partial_t f = D\partial_x^2 f + v (\partial_x h )(\partial_x f ) + \sqrt{f(1-f)/\bar{N}_T}\xi(x,t)\; \label{eq:dtf} \\
    \partial_t h = v + \lambda \partial_x^2 h + \frac{\nu}{2}(\partial_x h )^2  + \sqrt{2D_h}\eta(x,t) \;, \label{eq:dth}
\end{gather}
where $v$ is the velocity of the expanding colony and $\xi(x,t)$ is an independent Gaussian white noise process with zero mean.
Note that the important $v (\partial_x h )(\partial_x f )$ term in Eq.~\eqref{eq:dtf} can be interpreted as advection of $f$ controlled by the tilt $\partial_x h$ of the interface describing the frontier. 
It seems plausible that simpler frontier models apply to the diffusion-everywhere models of interest to us here.
As discussed below, this advective term leads to the focusing of genetic interfaces described in Fig.~\ref{fig:undul}.
In the neutral case $\epsilon=0$, we focus on now, and hence there is no selection advantage term $s f(1-f)$ in Eq.~\eqref{eq:dtf}. For identical reproduction rates $\mu$ between both species, when dilute, we expect that the equation for $h(x,t)$ is independent of $f(x,t)$.
As usual for such models with multiplicative noise, the noise term in Eq.~\eqref{eq:dtf} has to be interpreted in the Ito sense~\cite{Hallatschek2009}.

The dynamics of $h(x,t)$ is described by KPZ-dynamics Eq.~\eqref{eq:dth}~\cite{Kardar1986, Barabsi1995}, where $\eta(x,t)$ is a conventional Gaussian noise source. 
Although solving the fully coupled dynamics of Eqs.~\eqref{eq:dtf} and \eqref{eq:dth} for neutral frontier interfaces would be quite interesting, here, we provide only a scaling argument on how the width $w_{AB}$ at the frontier scales with $\bar{N}_T$.
For a simple flat front with $h(t) = v t$, $v (\partial_x h )(\partial_x f ) = 0$ and the frontier gentic fraction $f(x,t)$ evolves independently from $h(x,t)$.
In the long time limit of such flat front models, it was argued that the width should scale linearly with $\bar{N}_T$~\cite{Hallatschek2009}. 
However, this scaling, especially for large values of $\bar{N}_T$, predicts much larger genetic interface widths than observed in our numerical simulations, where the width at the frontier scales with $\log(\bar{N}_T)$. 

To better understand the geometric effects described above,  we first solve Eq.~\eqref{eq:dtf} the genetic fractional dynamics numerically by replacing Eq.~\eqref{eq:dth} with a simple cases of a stationary colony height profile, moving at velocity $v$ along $y$, with a step-like initial condition in the frontier genotype along $x$, i.e., $f(x) = \Theta(x_0-x)$ at $t=0$ and ignoring the genetic drift embodied in number fluctuations.
If the growing front is constantly tilted so that $h(x,t) = v t + a x$, the genetic fraction that solves Eq.~\eqref{eq:dtf} is then
\begin{equation}
\label{eq:flat_tilt_h}
    f(x,t) = \frac{1}{2}\left[ 1 + \text{erf} \left(\frac{ -(x-x_0+ a v t)}{2\sqrt{D t}} \right) \right].
\end{equation}
On the other hand, for a growing front of a parabolic shape $h(x) = b (x-x_0)^2/2$, we find
\begin{equation}
\label{eq:curved_h}
    f(x,t) = \frac{1}{2}\left[ 1 + \text{erf} \left(\frac{ -(x-x_0) \exp( v b t)}{\sqrt{2 D(\exp(2 b v t)-1)/(v b)}} \right) \right].
\end{equation}
Here, the initial step broadens faster as predicted by diffusion when $b<0$, i.e. for a front with a maxima at $x=x_0$. However, for a front with a minium $b>0$, Eq.~\eqref{eq:curved_h} predicts the finite width 
\begin{equation}
\label{eq:wAB_parab}
    w_{AB}=\sqrt{2 D/(v b)}
\end{equation} 
in the limit $t\rightarrow \infty$.
Both predictions from this highly simplified model of the frontier Eq.~\eqref{eq:dtf}, for the broadening and drift for tilted frontiers, and a focusing or de-focusing for curved interfaces, dependent on the sign of the curvature, agree qualitatively with the observed dynamics of the full two-dimensional dynamics shown in Fig.~\ref{fig:undul}, even though we neglect the number fluctuations that would give rise to undulating frontiers.

We now argue that the logarithmic scaling of the width of the genetic interface can be understood by first approximating the local curvature of a more general undulated frontier by a parabola and then using equation Eq.~\eqref{eq:wAB_parab} to predict the corresponding frontier genetic interface width.
The dynamics of a stochastic FKPP equation using Eq.~\eqref{eq:dth} develops according to an autonomous KPZ dynamics~\cite{Kardar1986, Barabsi1995, Moro2001}. 
The spectrum of the height fluctuation of KPZ interfaces follows $S(k) = C/k^\alpha$~\cite{Forster1977, Barabsi1995, Takeuchi2017} in the low $k$ region with $\alpha=-2$. 
A similar spectrum was found to describe the frontiers of colonies expanding according to a stochastic FKPP~\cite{Nesic2014}. Furthermore, spectra of different noise levels, i.e., carrying capacity $\bar{N}_T$, can be collapsed to a universal curve when the frequency axis $k$ is scaled by $\log(\bar{N}_T)$~\cite{Nesic2014}.
Therefore, when the local curvature at minima of a KPZ interface is approximated by a parabola with curvature $b$, we expect this curvature to scale with $1/\log^2(\bar{N}_T)$. 
Upon combining this scaling with the result for $w_{AB}$ at the frontier of a curved frontier in Eq.~\ref{eq:wAB_parab}, we arrive at $w_{AB}\propto \log(N_T)$, the scaling we find numerically for the full two-dimensional dynamics.

\textit{Conclusion - }
We studied the dynamics of the interface between two genotypes simultaneously expanding at the frontier of colonies generated via a stochastic FKPP dynamics. Our central finding is that, unlike in the bulk, where interspecies interactions control the genetic interface, the width $w_{AB}$ at the frontier is largely independent of these interactions and maintains a finite value at long times.
This width arises due to a focusing mechanism associated with the undulations of the colony's leading edge, which are inevitable in the presence of number fluctuations.

Specifically, we find that genetic interfaces with diffusion everywhere behind the front tend to drift towards local minima of the frontier height profile, where their broadening is arrested due to geometric focusing of growth. 
Via numerical simulations, we uncovered a logarithmic dependence of the interface width on the noise level of the stochastic FKPP equation, which is set by the local carrying capacity $\bar{N}_T$.
We conjecture that this scaling can be captured by an effectively one-dimensional frontier model (Eqs.~\eqref{eq:dtf} and \eqref{eq:dth}) where the dynamics of the local fraction is coupled to the colony height.
While simulation reveal a logarithmic scaling within our numerical accuracy, a rigorous theoretical derivation of this behavior is currently absent. Developing such a theory remains an important direction for future research.

These results highlight the fascinating interplay of geometric and stochastic effects at the leading edge of expanding populations. 
Embodied in a minimal model of a stochastic FKPP dynamics, they could be relevant not only for understanding the spatial structure of the range expansion of motile microbes, but also for understanding more generic expansions of invasive species into ecosystems, e.g., the Cane Toad invasion in Australia~\cite{EASTEAL1981, Shine2010}, or tumor growth~\cite{Friedl2011, Wirtz2011}.

\textit{Acknowledgment - }
We thank K.~Korolev, O.~Hallatschek, and M.~Kardar for enlightening discussions. This work was supported by the NSF thorugh the Harvard Materials Science and Engineering Center, through Grand DMR-2011754. J.B. thanks the German Research Foundation for financial support through the DFG Project BA 8210/1-1.

\clearpage
\appendix
  
\subsection{Appendix A: Bulk dynamics for mutualistic interactions}
\label{app:wedge}
\begin{figure}[t] 
    \includegraphics[width=0.99\linewidth]{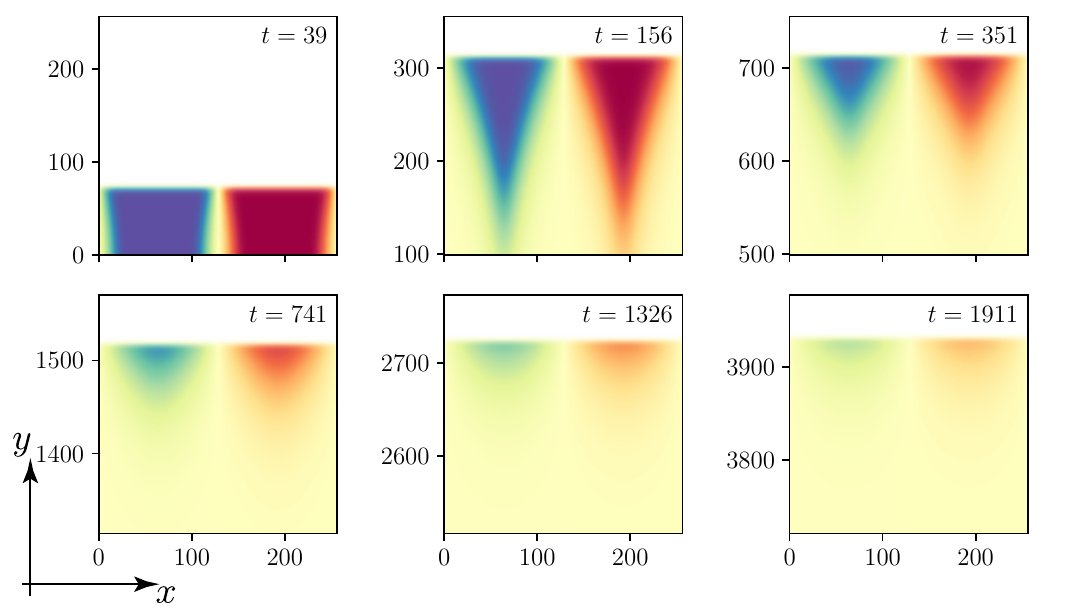}
    \caption{
    \textbf{Wedge at the frontier of mutualistic mixtures}:\label{fig:SI0} 
    Concentration profiles with deterministic dynamics with a frontier at six different times $t$ for mutualistic interactions $\epsilon =0.1$. 
    The blue and red genotypes in the bulk are replaced by a stable yellow mixed phase via a distinct pulled Fisher wave, but with slower diffusive mixing at the frontier itself.
    Periodic boundary conditions are employed in the x-direction, so that there is an interface both in the center and also at the boundary. Units: $ \lambda = \sqrt{D/\mu}$ (length), $\tau = 1/\mu$ (time); System-dimensions: $L_X = 256$, $L_Y = 4096$.
 } 
\end{figure}
For mutualistic interspecies interactions, a mixed state with both species present is the stable fixed point of the dynamics in the dense state. 
While each blue and red species at the frontier expands into unoccupied territory with the usual pulled Fisher wave, a second pulled wave follows, converting a domain of a single species into a well-mixed population. 
As discussed in the main text, the velocity of the second wave is always slower than the velocity of the first Fisher wave.
However, in the deterministic case, the two species at the frontier mix on a diffusive time scale.
The deterministic simulation in Fig.~\ref{fig:SI0} reveals these two time scales: Shortly after the initialization of the same step-like initial conditions as in the main text, the yellow mixed state establishes itself deep within the bulk (top left). In the remaining five parts of Fig.~\ref{fig:SI0}, we show the colony in the co-moving reference frame of the Fisher-wave. 
First, a wedge shape develops (boundary between blue/yellow and red/yellow), determined by the ratio of the two different velocities $v_\text{FKPP}=2\sqrt{D \mu}$ (colony growth into the unocupied territory) and $v_\text{mut}=\sqrt{2 D \mu \epsilon}$ (mixed 50-50 concentration, established at the expense of the pure genotype). On the slower diffusive timescale, the domains of single species at the frontier gradually blur into the yellow well-mixed state.

\subsection{Appendix B: Measuring the $AB$-interface width $w_{AB}$}
\begin{figure}[t] 
    \includegraphics[width=0.99\linewidth]{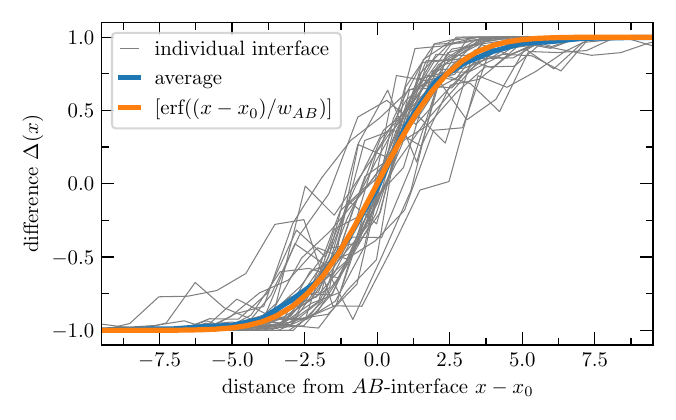}
    \caption{
    \textbf{Profiles of the difference $\Delta(x) = (N_A-N_B)/N_T$}\label{fig:SI5} 
    across an $AB$-interface in $x$-direction at the frontier of a colony. The frontier is defined as the height where $N_T = \bar{N}_T/2$ in lattice simulations  with $\bar{N}_T=100$. Profiles of individual runs are shown by thin, gray lines, their average is shown by the blue line. 
    For every individual profile the width $w_{AB}$ and interface position $x_0$ were obtained via fitting an error function. An error function profile with the average width $w_{AB}\approx 2.96$ is shown in orange, and provides a reasonable fit to the average shown in blue.
 } 
\end{figure}
In our stochastic agend-based simulations, we define the frontier of the colony at a given time along the $x$-direction, as the height in $y$-direction at which $N_T = \bar{N}_T/2$. Across this frontier, we measure the difference $\Delta(x) = (N_A - N_B)/N_T$ as a function of $x$. Along this direction, we find two $AB$-interfaces, i.e., one from $A$ to $B$ and another from $B$ to $A$, due to our periodic boundary conditions. Up to a sign, these interfaces are identical, and we obtain their positions $x_0$ and widths $w_{AB}$ from fits of an error function for each of them. In Fig.~\ref{fig:SI5}, we show individual of such profiles centered around the interface position $x_0$ by the thin gray lines. 
We also show the average profile, averaged over all the individual runs (blue line), and an error function profile with the corresponding average width $w_{AB}$ in orange.
Close to the center of the interface, the error function seems to describe the average profile well. In the tails, the average profile seems to decay a bit more slowly than predicted by the error function.

\subsection{Appendix C: $AB$-interface widths for FKPP-waves with demographic noise}
\label{app:demograph}
\begin{figure}[t!] 
    \includegraphics[width=0.99\linewidth]{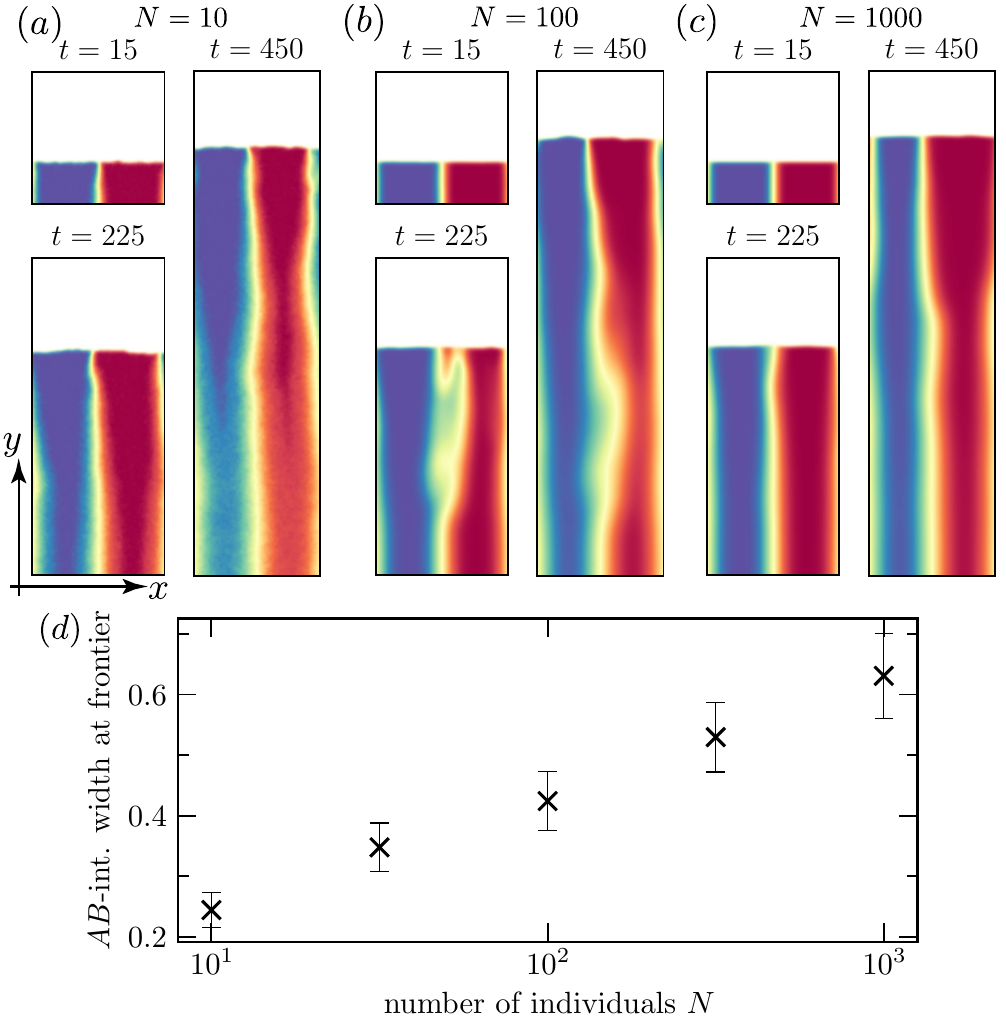}
    \caption{
    \textbf{Demographic noise in binary range expansion}:\label{fig:SI2} 
    (a-c) Concentration fields at three different times $t$ for neutral interactions and demographic noise $N=10,100,1000$.
    (d) $AB$-interface width measured along the x-direction at time $t=450$ for five independent runs, where we excluded rare simulations that showed long-lived branching events, i.e., those that survived long enough to be present at the frontier when $t=450$. A more short-lived branching event is shown in (b) for $t=225$.
    Parameters and initial conditions are identical to Fig.~\ref{fig:FKPP}.
    } 
\end{figure}
In the main part of this work, we used birth and death processes on a hexagonal lattice (with centers that form a triangular lattice) to generate a stochastic version of the dynamics embodied in the deterministic FKPP equation. In this Appendix, we describe simulation evidence that the same logarithmic scaling of the $AB$-interface width can be found when solving the FKPP equation with simple demographic noise model. We solve the stochastic partial differential equations
\begin{equation}
    \label{eq:FKPP_stochastic}
    \partial_t c_i = D \nabla^2 c_i  + \mu c_i  (1  - c_A - c_B + \epsilon  c_j) +\sqrt{c_i/N} \xi_i(x,y,t)\, ,
\end{equation} 
with $i=A,B$, $j\neq i$ and white Gaussian noise processes $\langle \xi_i(x,y,t) \rangle=0$, and $\langle \xi_i(x,y,t)\xi_i(x',y',t') \rangle=\delta(x-x')\delta(y-y')\delta(t-t')\delta_{ij}$ using the splitting scheme introduced by \cite{Pechenik1999} and further optimized by \cite{Dornic2005}. 
In Fig.~\ref{fig:SI2}(a-c), we show three typical results for $N=10,100,1000$, together with the interface width $w_{AB}$, showing the same logarithmic dependency on $N$ as we found with the average deme size $\bar{N}_T$ for the stochastic simulations in Fig.~\ref{fig:width_vs_N}.

\subsection{Appendix D: Long-lived splitting of genetic interfaces}
\begin{figure}[t!] 
    \includegraphics[width=0.99\linewidth]{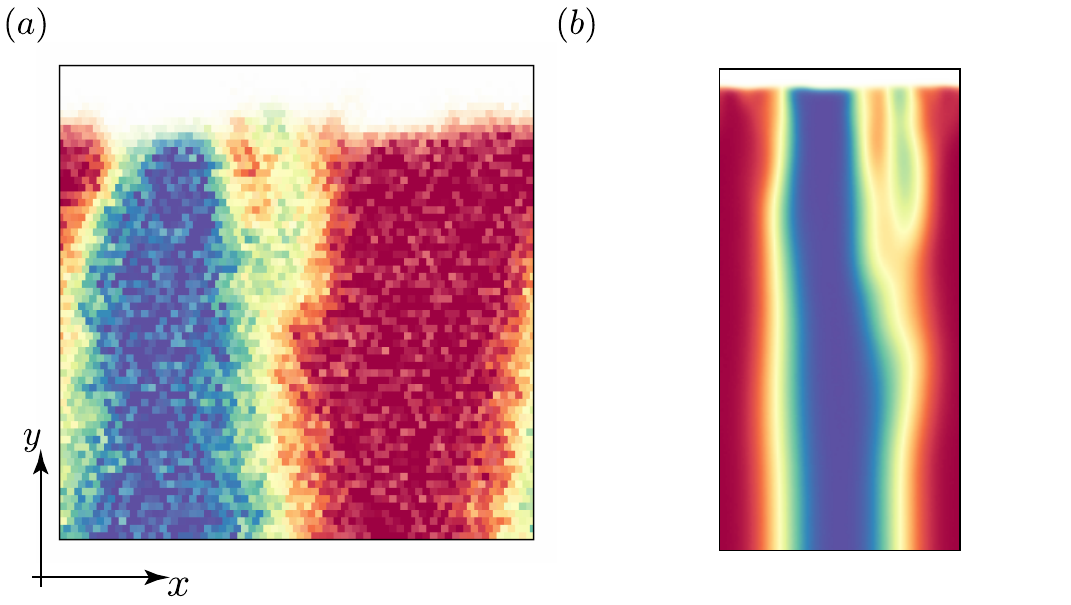}
    \caption{
    \textbf{Splitting of a genetic interface}\label{fig:SI3} 
    in three genetic interfaces for neutral mixtures:
    (a) for a lattice simulation $\bar{N}_T=100$ at time $t=500$; same parameters as in Fig.~\ref{fig:width_vs_N}
    (b) for a numerical simulation of the stochastic FKPP wave with $N=100$ at time $t=450$; same parameters as in Fig.~\ref{fig:SI2}
    } 
\end{figure}
In our measurements of the genetic interface width, we excluded relatively rare simulations where one genetic interface was split into long-lived three genetic interfaces, indicated schematically by $AAA|BBB \rightarrow AA|B|A|BB$, thus causing our fit of the genetic fraction at the frontier to an error function profile to fail. In Fig.~\ref{fig:SI3}, we show examples of such splitting events. For the numerically obtained results of the genetic interface width at the frontier, shown in Fig.~\ref{fig:width_vs_N}(c) of the main text, we excluded between one and three different runs from the 20 independent runs that were averaged.

\subsection{Appendix E: A flat frontier lattice model with number fluctuations and diffusion in the bulk}
\label{app:flat_front_broad}
We now illustrate the importance of frontier undulations by studying the $AB$-interface at frontiers with number fluctuations, but nevertheless are forced to stay flat.  
To do this, we introduce a version of our stochastic lattice model that decouples the diffusive steps from the expansion of the colony at the frontier.
We use the same triangular lattice of cells with one of the three principal nearest neighbor directions aligned with the $x$-axis. In every deme, the birth- and death-processes Eqs.~\eqref{eq:birth} and \eqref{eq:death} in the main textoccur. Away from the frontier, every individual can jump at a rate $\rho$ to one of the six neighboring demes, as in the previous model. 
However, at the frontier, diffusive exchanges only take place between the four already occupied neighbors, while jumps into the two neighboring sites that belong to the next row in $y$-direction are prohibited.
This new row is populated instead after a generational lifetime $T_\text{gen}$. After this time, every current individual at the frontier reproduces with a rate $\mu_\text{off}$. These offspring individuals are positioned with an equal likelihood to one of two demes of the next generation, thus ensuring a flat front. It is as if flatness were enforced by an extremely large line tension between occupied and unoccupied territory in this $\epsilon=0$ neutral model.

\begin{figure}[!b] 
    \includegraphics[width=0.99\linewidth]{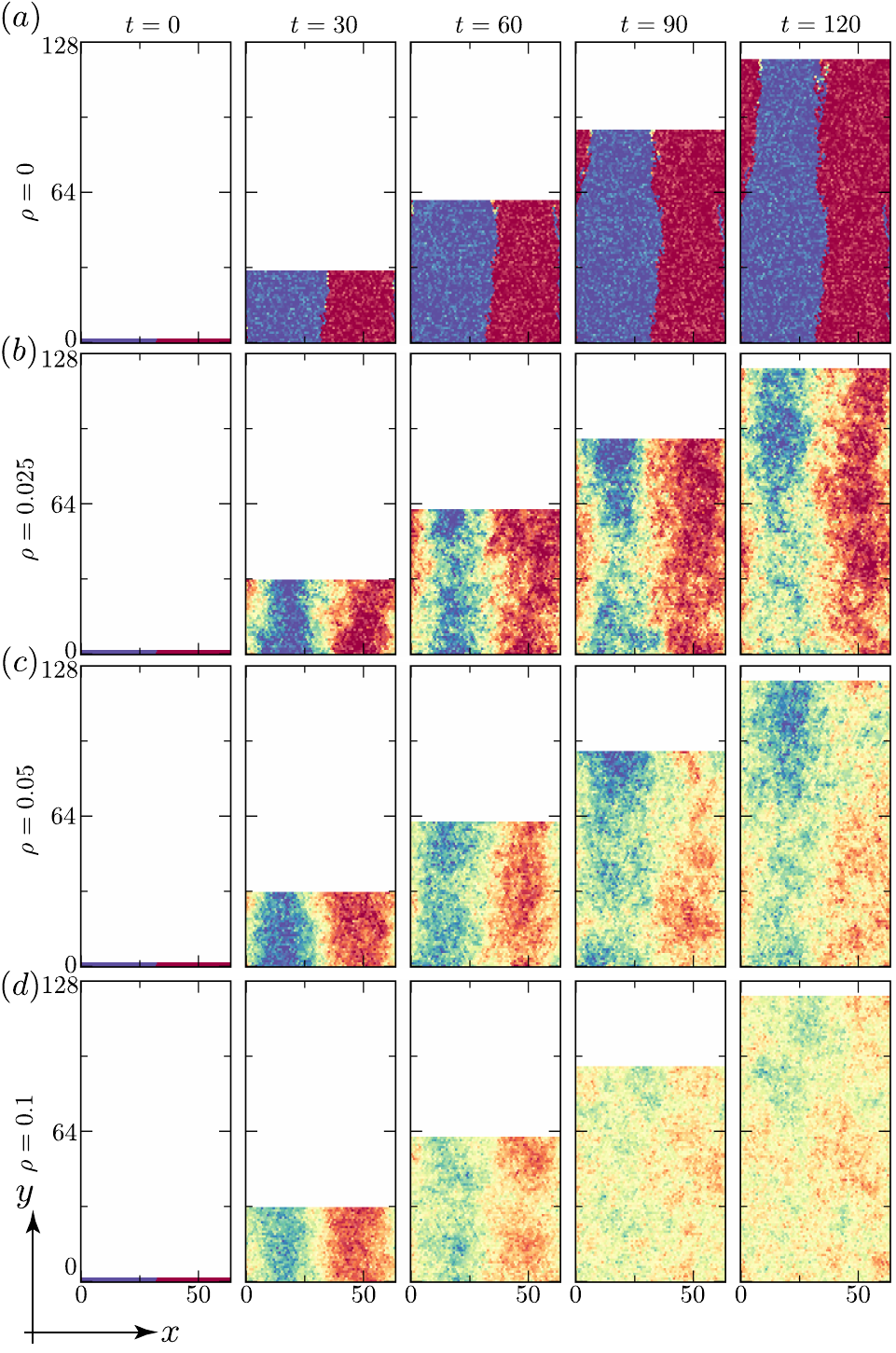}
    \caption{
    \textbf{Flat fronts with number fluctuations and bulk diffusion}:\label{fig:SI1} 
    Typical lattice configurations of the amended lattice model that prevents undulations of the front at five time points (for different jump rates $\rho=0, 0.025, 0.05, 0.1$ (a-d)), with $\epsilon=0$.
    Within the bulk, the same rules for birth, death and diffusion apply as in the model used in the main text. However, at the frontier, diffusive jumps that would lead to individuals in the next row are prohibited.  
    Instead, after a generational time $T_\text{gen}$, individuals at the frontier can reproduce with rate $\mu_\text{off}$ and their offspring is positioned in one of the two neighboring lattice sites in the row above.
    Parameters: $\mu=0.1$, $\lambda_\text{self}=\lambda_\text{cross}=0.001$ (neutral, $\bar{N}_T=100$), $\mu_\text{off}=0.1$, $T_\text{gen}=100$, $\rho=0, 0.025, 0.05, 0.1$.
    } 
\end{figure}

In Fig.~\ref{fig:SI1}(a-d), we show typical results for simulations with $\rho=0, 0.025, 0.05, 0.1$, where initially, the first generation is fully occupied by $A$/$B$ individuals in the left/right half. Only in the case of $\rho=0$ does the $AB$-interface width between the two domains survive over all generations, see Fig.~\ref{fig:SI1}(a).
Otherwise, bulk diffusion destroys the genetic interface both in the bulk and at the frontier.
Every deme, once initialized at the beginning of the lifetime of every generation, is entirely independent from the others. Thus, similar to the Moran process, fixation can be reached within every deme during the generational lifetime $T_\text{gen}$. 
Thus, we expect every deme to be populated only by either $A$ or $B$ in this 1+1 dimensional model in the limit of large generational lifetimes $T_\text{gen}$.
However, when $\rho>0$, diffusion can broaden the $AB$-interface even within one generation. Across multiple generations, diffusion continues to broaden initially sharp interfaces, as shown in Fig.~\ref{fig:SI1}(b-d). In this 2+1 dimensional flat front model (two dimensions of space and one of time), we find much broader interfaces, or even completely mixed fronts, compared to the 2+1 dimensional model with undulating frontiers.
Understanding the differences in the long time dynamics between flat 2+1 dimensional models and 1+1 dimensional models, as studied in Ref.~\cite{Hallatschek2009}, is left for future work.

\clearpage


\begin{thebibliography}{36}%
\makeatletter
\providecommand \@ifxundefined [1]{%
 \@ifx{#1\undefined}
}%
\providecommand \@ifnum [1]{%
 \ifnum #1\expandafter \@firstoftwo
 \else \expandafter \@secondoftwo
 \fi
}%
\providecommand \@ifx [1]{%
 \ifx #1\expandafter \@firstoftwo
 \else \expandafter \@secondoftwo
 \fi
}%
\providecommand \natexlab [1]{#1}%
\providecommand \enquote  [1]{``#1''}%
\providecommand \bibnamefont  [1]{#1}%
\providecommand \bibfnamefont [1]{#1}%
\providecommand \citenamefont [1]{#1}%
\providecommand \href@noop [0]{\@secondoftwo}%
\providecommand \href [0]{\begingroup \@sanitize@url \@href}%
\providecommand \@href[1]{\@@startlink{#1}\@@href}%
\providecommand \@@href[1]{\endgroup#1\@@endlink}%
\providecommand \@sanitize@url [0]{\catcode `\\12\catcode `\$12\catcode
  `\&12\catcode `\#12\catcode `\^12\catcode `\_12\catcode `\%12\relax}%
\providecommand \@@startlink[1]{}%
\providecommand \@@endlink[0]{}%
\providecommand \url  [0]{\begingroup\@sanitize@url \@url }%
\providecommand \@url [1]{\endgroup\@href {#1}{\urlprefix }}%
\providecommand \urlprefix  [0]{URL }%
\providecommand \Eprint [0]{\href }%
\providecommand \doibase [0]{https://doi.org/}%
\providecommand \selectlanguage [0]{\@gobble}%
\providecommand \bibinfo  [0]{\@secondoftwo}%
\providecommand \bibfield  [0]{\@secondoftwo}%
\providecommand \translation [1]{[#1]}%
\providecommand \BibitemOpen [0]{}%
\providecommand \bibitemStop [0]{}%
\providecommand \bibitemNoStop [0]{.\EOS\space}%
\providecommand \EOS [0]{\spacefactor3000\relax}%
\providecommand \BibitemShut  [1]{\csname bibitem#1\endcsname}%
\let\auto@bib@innerbib\@empty
\bibitem [{\citenamefont {Fisher}(1937)}]{Fisher1937}%
  \BibitemOpen
  \bibfield  {author} {\bibinfo {author} {\bibfnamefont {R.~A.}\ \bibnamefont
  {Fisher}},\ }\bibfield  {title} {\bibinfo {title} {The wave of advance of
  advantageous genes},\ }\href
  {https://doi.org/10.1111/j.1469-1809.1937.tb02153.x} {\bibfield  {journal}
  {\bibinfo  {journal} {Annals of Eugenics}\ }\textbf {\bibinfo {volume} {7}},\
  \bibinfo {pages} {355–369} (\bibinfo {year} {1937})}\BibitemShut {NoStop}%
\bibitem [{\citenamefont {Kolmogorov}\ \emph {et~al.}(1937)\citenamefont
  {Kolmogorov}, \citenamefont {Petrovsky},\ and\ \citenamefont
  {Piscounov}}]{Kolmogorov1937}%
  \BibitemOpen
  \bibfield  {author} {\bibinfo {author} {\bibfnamefont {A.~N.}\ \bibnamefont
  {Kolmogorov}}, \bibinfo {author} {\bibfnamefont {I.~G.}\ \bibnamefont
  {Petrovsky}},\ and\ \bibinfo {author} {\bibfnamefont {N.~S.}\ \bibnamefont
  {Piscounov}},\ }\bibfield  {title} {\bibinfo {title} {Study of the diffusion
  equation with growth of the quantity of matter and its application to a
  biological problem},\ }\href@noop {} {\bibfield  {journal} {\bibinfo
  {journal} {Bull. Moscow Univ. Math. Mech.}\ }\textbf {\bibinfo {volume}
  {1}},\ \bibinfo {pages} {1} (\bibinfo {year} {1937})}\BibitemShut {NoStop}%
\bibitem [{\citenamefont {Murray}(2002)}]{Murray2002}%
  \BibitemOpen
  \bibfield  {author} {\bibinfo {author} {\bibfnamefont {J.~D.}\ \bibnamefont
  {Murray}},\ }\href@noop {} {\emph {\bibinfo {title} {Mathematical
  Biology}}},\ \bibinfo {edition} {3rd}\ ed.,\ edited by\ \bibinfo {editor}
  {\bibfnamefont {J.~D.}\ \bibnamefont {Murray}},\ Interdisciplinary applied
  mathematics\ (\bibinfo  {publisher} {Springer},\ \bibinfo {address} {New
  York, NY},\ \bibinfo {year} {2002})\BibitemShut {NoStop}%
\bibitem [{\citenamefont {Hallatschek}\ \emph {et~al.}(2023)\citenamefont
  {Hallatschek}, \citenamefont {Datta}, \citenamefont {Drescher}, \citenamefont
  {Dunkel}, \citenamefont {Elgeti}, \citenamefont {Waclaw},\ and\ \citenamefont
  {Wingreen}}]{Hallatschek2023}%
  \BibitemOpen
  \bibfield  {author} {\bibinfo {author} {\bibfnamefont {O.}~\bibnamefont
  {Hallatschek}}, \bibinfo {author} {\bibfnamefont {S.~S.}\ \bibnamefont
  {Datta}}, \bibinfo {author} {\bibfnamefont {K.}~\bibnamefont {Drescher}},
  \bibinfo {author} {\bibfnamefont {J.}~\bibnamefont {Dunkel}}, \bibinfo
  {author} {\bibfnamefont {J.}~\bibnamefont {Elgeti}}, \bibinfo {author}
  {\bibfnamefont {B.}~\bibnamefont {Waclaw}},\ and\ \bibinfo {author}
  {\bibfnamefont {N.~S.}\ \bibnamefont {Wingreen}},\ }\bibfield  {title}
  {\bibinfo {title} {Proliferating active matter},\ }\href
  {https://doi.org/10.1038/s42254-023-00593-0} {\bibfield  {journal} {\bibinfo
  {journal} {Nature Reviews Physics}\ }\textbf {\bibinfo {volume} {5}},\
  \bibinfo {pages} {407–419} (\bibinfo {year} {2023})}\BibitemShut {NoStop}%
\bibitem [{\citenamefont {Farrell}\ \emph {et~al.}(2013)\citenamefont
  {Farrell}, \citenamefont {Hallatschek}, \citenamefont {Marenduzzo},\ and\
  \citenamefont {Waclaw}}]{Farrell2013}%
  \BibitemOpen
  \bibfield  {author} {\bibinfo {author} {\bibfnamefont {F.~D.~C.}\
  \bibnamefont {Farrell}}, \bibinfo {author} {\bibfnamefont {O.}~\bibnamefont
  {Hallatschek}}, \bibinfo {author} {\bibfnamefont {D.}~\bibnamefont
  {Marenduzzo}},\ and\ \bibinfo {author} {\bibfnamefont {B.}~\bibnamefont
  {Waclaw}},\ }\bibfield  {title} {\bibinfo {title} {Mechanically driven growth
  of quasi-two-dimensional microbial colonies},\ }\bibfield  {journal}
  {\bibinfo  {journal} {Physical Review Letters}\ }\textbf {\bibinfo {volume}
  {111}},\ \href {https://doi.org/10.1103/physrevlett.111.168101}
  {10.1103/physrevlett.111.168101} (\bibinfo {year} {2013})\BibitemShut
  {NoStop}%
\bibitem [{\citenamefont {Giometto}\ \emph {et~al.}(2018)\citenamefont
  {Giometto}, \citenamefont {Nelson},\ and\ \citenamefont
  {Murray}}]{Giometto2018}%
  \BibitemOpen
  \bibfield  {author} {\bibinfo {author} {\bibfnamefont {A.}~\bibnamefont
  {Giometto}}, \bibinfo {author} {\bibfnamefont {D.~R.}\ \bibnamefont
  {Nelson}},\ and\ \bibinfo {author} {\bibfnamefont {A.~W.}\ \bibnamefont
  {Murray}},\ }\bibfield  {title} {\bibinfo {title} {Physical interactions
  reduce the power of natural selection in growing yeast colonies},\ }\href
  {https://doi.org/10.1073/pnas.1809587115} {\bibfield  {journal} {\bibinfo
  {journal} {Proceedings of the National Academy of Sciences}\ }\textbf
  {\bibinfo {volume} {115}},\ \bibinfo {pages} {11448–11453} (\bibinfo {year}
  {2018})}\BibitemShut {NoStop}%
\bibitem [{\citenamefont {Martínez-Calvo}\ \emph {et~al.}(2022)\citenamefont
  {Martínez-Calvo}, \citenamefont {Bhattacharjee}, \citenamefont {Bay},
  \citenamefont {Luu}, \citenamefont {Hancock}, \citenamefont {Wingreen},\ and\
  \citenamefont {Datta}}]{MartnezCalvo2022}%
  \BibitemOpen
  \bibfield  {author} {\bibinfo {author} {\bibfnamefont {A.}~\bibnamefont
  {Martínez-Calvo}}, \bibinfo {author} {\bibfnamefont {T.}~\bibnamefont
  {Bhattacharjee}}, \bibinfo {author} {\bibfnamefont {R.~K.}\ \bibnamefont
  {Bay}}, \bibinfo {author} {\bibfnamefont {H.~N.}\ \bibnamefont {Luu}},
  \bibinfo {author} {\bibfnamefont {A.~M.}\ \bibnamefont {Hancock}}, \bibinfo
  {author} {\bibfnamefont {N.~S.}\ \bibnamefont {Wingreen}},\ and\ \bibinfo
  {author} {\bibfnamefont {S.~S.}\ \bibnamefont {Datta}},\ }\bibfield  {title}
  {\bibinfo {title} {Morphological instability and roughening of growing 3d
  bacterial colonies},\ }\bibfield  {journal} {\bibinfo  {journal} {Proceedings
  of the National Academy of Sciences}\ }\textbf {\bibinfo {volume} {119}},\
  \href {https://doi.org/10.1073/pnas.2208019119} {10.1073/pnas.2208019119}
  (\bibinfo {year} {2022})\BibitemShut {NoStop}%
\bibitem [{\citenamefont {Hallatschek}\ \emph {et~al.}(2007)\citenamefont
  {Hallatschek}, \citenamefont {Hersen}, \citenamefont {Ramanathan},\ and\
  \citenamefont {Nelson}}]{Hallatschek2007}%
  \BibitemOpen
  \bibfield  {author} {\bibinfo {author} {\bibfnamefont {O.}~\bibnamefont
  {Hallatschek}}, \bibinfo {author} {\bibfnamefont {P.}~\bibnamefont {Hersen}},
  \bibinfo {author} {\bibfnamefont {S.}~\bibnamefont {Ramanathan}},\ and\
  \bibinfo {author} {\bibfnamefont {D.~R.}\ \bibnamefont {Nelson}},\ }\bibfield
   {title} {\bibinfo {title} {Genetic drift at expanding frontiers promotes
  gene segregation},\ }\href {https://doi.org/10.1073/pnas.0710150104}
  {\bibfield  {journal} {\bibinfo  {journal} {Proceedings of the National
  Academy of Sciences}\ }\textbf {\bibinfo {volume} {104}},\ \bibinfo {pages}
  {19926–19930} (\bibinfo {year} {2007})}\BibitemShut {NoStop}%
\bibitem [{\citenamefont {Saito}\ and\ \citenamefont
  {M\"{u}ller-Krumbhaar}(1995)}]{Saito1995}%
  \BibitemOpen
  \bibfield  {author} {\bibinfo {author} {\bibfnamefont {Y.}~\bibnamefont
  {Saito}}\ and\ \bibinfo {author} {\bibfnamefont {H.}~\bibnamefont
  {M\"{u}ller-Krumbhaar}},\ }\bibfield  {title} {\bibinfo {title} {Critical
  phenomena in morphology transitions of growth models with competition},\
  }\href {https://doi.org/10.1103/physrevlett.74.4325} {\bibfield  {journal}
  {\bibinfo  {journal} {Physical Review Letters}\ }\textbf {\bibinfo {volume}
  {74}},\ \bibinfo {pages} {4325–4328} (\bibinfo {year} {1995})}\BibitemShut
  {NoStop}%
\bibitem [{\citenamefont {Hallatschek}\ and\ \citenamefont
  {Nelson}(2010)}]{Hallatschek2010}%
  \BibitemOpen
  \bibfield  {author} {\bibinfo {author} {\bibfnamefont {O.}~\bibnamefont
  {Hallatschek}}\ and\ \bibinfo {author} {\bibfnamefont {D.~R.}\ \bibnamefont
  {Nelson}},\ }\bibfield  {title} {\bibinfo {title} {Life at the front of an
  expanding population},\ }\href
  {https://doi.org/10.1111/j.1558-5646.2009.00809.x} {\bibfield  {journal}
  {\bibinfo  {journal} {Evolution}\ }\textbf {\bibinfo {volume} {64}},\
  \bibinfo {pages} {193–206} (\bibinfo {year} {2010})}\BibitemShut {NoStop}%
\bibitem [{\citenamefont {Pigolotti}\ \emph {et~al.}(2013)\citenamefont
  {Pigolotti}, \citenamefont {Benzi}, \citenamefont {Perlekar}, \citenamefont
  {Jensen}, \citenamefont {Toschi},\ and\ \citenamefont
  {Nelson}}]{Pigolotti2013}%
  \BibitemOpen
  \bibfield  {author} {\bibinfo {author} {\bibfnamefont {S.}~\bibnamefont
  {Pigolotti}}, \bibinfo {author} {\bibfnamefont {R.}~\bibnamefont {Benzi}},
  \bibinfo {author} {\bibfnamefont {P.}~\bibnamefont {Perlekar}}, \bibinfo
  {author} {\bibfnamefont {M.}~\bibnamefont {Jensen}}, \bibinfo {author}
  {\bibfnamefont {F.}~\bibnamefont {Toschi}},\ and\ \bibinfo {author}
  {\bibfnamefont {D.}~\bibnamefont {Nelson}},\ }\bibfield  {title} {\bibinfo
  {title} {Growth, competition and cooperation in spatial population
  genetics},\ }\href {https://doi.org/10.1016/j.tpb.2012.12.002} {\bibfield
  {journal} {\bibinfo  {journal} {Theoretical Population Biology}\ }\textbf
  {\bibinfo {volume} {84}},\ \bibinfo {pages} {72–86} (\bibinfo {year}
  {2013})}\BibitemShut {NoStop}%
\bibitem [{\citenamefont {van Saarloos}(2003)}]{VANSAARLOOS2003}%
  \BibitemOpen
  \bibfield  {author} {\bibinfo {author} {\bibfnamefont {W.}~\bibnamefont {van
  Saarloos}},\ }\bibfield  {title} {\bibinfo {title} {Front propagation into
  unstable states},\ }\href {https://doi.org/10.1016/j.physrep.2003.08.001}
  {\bibfield  {journal} {\bibinfo  {journal} {Physics Reports}\ }\textbf
  {\bibinfo {volume} {386}},\ \bibinfo {pages} {29–222} (\bibinfo {year}
  {2003})}\BibitemShut {NoStop}%
\bibitem [{\citenamefont {Doering}\ \emph {et~al.}(2003)\citenamefont
  {Doering}, \citenamefont {Mueller},\ and\ \citenamefont
  {Smereka}}]{Doering2003}%
  \BibitemOpen
  \bibfield  {author} {\bibinfo {author} {\bibfnamefont {C.~R.}\ \bibnamefont
  {Doering}}, \bibinfo {author} {\bibfnamefont {C.}~\bibnamefont {Mueller}},\
  and\ \bibinfo {author} {\bibfnamefont {P.}~\bibnamefont {Smereka}},\
  }\bibfield  {title} {\bibinfo {title} {Interacting particles, the stochastic
  {F}isher–{K}olmogorov–{P}etrovsky–{P}iscounov equation, and duality},\
  }\href {https://doi.org/10.1016/s0378-4371(03)00203-6} {\bibfield  {journal}
  {\bibinfo  {journal} {Physica A: Statistical Mechanics and its Applications}\
  }\textbf {\bibinfo {volume} {325}},\ \bibinfo {pages} {243–259} (\bibinfo
  {year} {2003})}\BibitemShut {NoStop}%
\bibitem [{\citenamefont {Gillespie}(1977)}]{gillespie1977exact}%
  \BibitemOpen
  \bibfield  {author} {\bibinfo {author} {\bibfnamefont {D.~T.}\ \bibnamefont
  {Gillespie}},\ }\bibfield  {title} {\bibinfo {title} {Exact stochastic
  simulation of coupled chemical reactions},\ }\href
  {https://doi.org/10.1021/j100540a008} {\bibfield  {journal} {\bibinfo
  {journal} {The Journal of Physical Chemistry}\ }\textbf {\bibinfo {volume}
  {81}},\ \bibinfo {pages} {2340} (\bibinfo {year} {1977})}\BibitemShut
  {NoStop}%
\bibitem [{\citenamefont {Brunet}\ and\ \citenamefont
  {Derrida}(1997)}]{Brunet1997}%
  \BibitemOpen
  \bibfield  {author} {\bibinfo {author} {\bibfnamefont {E.}~\bibnamefont
  {Brunet}}\ and\ \bibinfo {author} {\bibfnamefont {B.}~\bibnamefont
  {Derrida}},\ }\bibfield  {title} {\bibinfo {title} {Shift in the velocity of
  a front due to a cutoff},\ }\href {https://doi.org/10.1103/physreve.56.2597}
  {\bibfield  {journal} {\bibinfo  {journal} {Physical Review E}\ }\textbf
  {\bibinfo {volume} {56}},\ \bibinfo {pages} {2597–2604} (\bibinfo {year}
  {1997})}\BibitemShut {NoStop}%
\bibitem [{\citenamefont {Brunet}\ and\ \citenamefont
  {Derrida}(2001)}]{Brunet2001}%
  \BibitemOpen
  \bibfield  {author} {\bibinfo {author} {\bibfnamefont {E.}~\bibnamefont
  {Brunet}}\ and\ \bibinfo {author} {\bibfnamefont {B.}~\bibnamefont
  {Derrida}},\ }\bibfield  {title} {\bibinfo {title} {Effect of microscopic
  noise on front propagation},\ }\href
  {https://doi.org/10.1023/a:1004875804376} {\bibfield  {journal} {\bibinfo
  {journal} {Journal of Statistical Physics}\ }\textbf {\bibinfo {volume}
  {103}},\ \bibinfo {pages} {269–282} (\bibinfo {year} {2001})}\BibitemShut
  {NoStop}%
\bibitem [{\citenamefont {Panja}(2004)}]{Panja2004}%
  \BibitemOpen
  \bibfield  {author} {\bibinfo {author} {\bibfnamefont {D.}~\bibnamefont
  {Panja}},\ }\bibfield  {title} {\bibinfo {title} {Effects of fluctuations on
  propagating fronts},\ }\href {https://doi.org/10.1016/j.physrep.2003.12.001}
  {\bibfield  {journal} {\bibinfo  {journal} {Physics Reports}\ }\textbf
  {\bibinfo {volume} {393}},\ \bibinfo {pages} {87–174} (\bibinfo {year}
  {2004})}\BibitemShut {NoStop}%
\bibitem [{\citenamefont {Drossel}\ and\ \citenamefont
  {Kardar}(2000)}]{Drossel2000}%
  \BibitemOpen
  \bibfield  {author} {\bibinfo {author} {\bibfnamefont {B.}~\bibnamefont
  {Drossel}}\ and\ \bibinfo {author} {\bibfnamefont {M.}~\bibnamefont
  {Kardar}},\ }\bibfield  {title} {\bibinfo {title} {Phase ordering and
  roughening on growing films},\ }\href
  {https://doi.org/10.1103/physrevlett.85.614} {\bibfield  {journal} {\bibinfo
  {journal} {Physical Review Letters}\ }\textbf {\bibinfo {volume} {85}},\
  \bibinfo {pages} {614–617} (\bibinfo {year} {2000})}\BibitemShut {NoStop}%
\bibitem [{\citenamefont {Chu}\ \emph {et~al.}(2019)\citenamefont {Chu},
  \citenamefont {Kardar}, \citenamefont {Nelson},\ and\ \citenamefont
  {Beller}}]{Chu2019}%
  \BibitemOpen
  \bibfield  {author} {\bibinfo {author} {\bibfnamefont {S.}~\bibnamefont
  {Chu}}, \bibinfo {author} {\bibfnamefont {M.}~\bibnamefont {Kardar}},
  \bibinfo {author} {\bibfnamefont {D.~R.}\ \bibnamefont {Nelson}},\ and\
  \bibinfo {author} {\bibfnamefont {D.~A.}\ \bibnamefont {Beller}},\ }\bibfield
   {title} {\bibinfo {title} {Evolution in range expansions with competition at
  rough boundaries},\ }\href {https://doi.org/10.1016/j.jtbi.2019.06.018}
  {\bibfield  {journal} {\bibinfo  {journal} {Journal of Theoretical Biology}\
  }\textbf {\bibinfo {volume} {478}},\ \bibinfo {pages} {153–160} (\bibinfo
  {year} {2019})}\BibitemShut {NoStop}%
\bibitem [{\citenamefont {Korolev}\ \emph {et~al.}(2010)\citenamefont
  {Korolev}, \citenamefont {Avlund}, \citenamefont {Hallatschek},\ and\
  \citenamefont {Nelson}}]{Korolev2010}%
  \BibitemOpen
  \bibfield  {author} {\bibinfo {author} {\bibfnamefont {K.~S.}\ \bibnamefont
  {Korolev}}, \bibinfo {author} {\bibfnamefont {M.}~\bibnamefont {Avlund}},
  \bibinfo {author} {\bibfnamefont {O.}~\bibnamefont {Hallatschek}},\ and\
  \bibinfo {author} {\bibfnamefont {D.~R.}\ \bibnamefont {Nelson}},\ }\bibfield
   {title} {\bibinfo {title} {Genetic demixing and evolution in linear stepping
  stone models},\ }\href {https://doi.org/10.1103/revmodphys.82.1691}
  {\bibfield  {journal} {\bibinfo  {journal} {Reviews of Modern Physics}\
  }\textbf {\bibinfo {volume} {82}},\ \bibinfo {pages} {1691–1718} (\bibinfo
  {year} {2010})}\BibitemShut {NoStop}%
\bibitem [{\citenamefont {Horowitz}\ and\ \citenamefont
  {Kardar}(2019)}]{Horowitz2019}%
  \BibitemOpen
  \bibfield  {author} {\bibinfo {author} {\bibfnamefont {J.~M.}\ \bibnamefont
  {Horowitz}}\ and\ \bibinfo {author} {\bibfnamefont {M.}~\bibnamefont
  {Kardar}},\ }\bibfield  {title} {\bibinfo {title} {Bacterial range expansions
  on a growing front: Roughness, fixation, and directed percolation},\
  }\bibfield  {journal} {\bibinfo  {journal} {Physical Review E}\ }\textbf
  {\bibinfo {volume} {99}},\ \href {https://doi.org/10.1103/physreve.99.042134}
  {10.1103/physreve.99.042134} (\bibinfo {year} {2019})\BibitemShut {NoStop}%
\bibitem [{\citenamefont {Swartz}\ \emph {et~al.}(2023)\citenamefont {Swartz},
  \citenamefont {Lee}, \citenamefont {Kardar},\ and\ \citenamefont
  {Korolev}}]{Swartz2023}%
  \BibitemOpen
  \bibfield  {author} {\bibinfo {author} {\bibfnamefont {D.~W.}\ \bibnamefont
  {Swartz}}, \bibinfo {author} {\bibfnamefont {H.}~\bibnamefont {Lee}},
  \bibinfo {author} {\bibfnamefont {M.}~\bibnamefont {Kardar}},\ and\ \bibinfo
  {author} {\bibfnamefont {K.~S.}\ \bibnamefont {Korolev}},\ }\bibfield
  {title} {\bibinfo {title} {Interplay between morphology and competition in
  two-dimensional colony expansion},\ }\bibfield  {journal} {\bibinfo
  {journal} {Physical Review E}\ }\textbf {\bibinfo {volume} {108}},\ \href
  {https://doi.org/10.1103/physreve.108.l032301} {10.1103/physreve.108.l032301}
  (\bibinfo {year} {2023})\BibitemShut {NoStop}%
\bibitem [{\citenamefont {Swartz}\ \emph {et~al.}(2024)\citenamefont {Swartz},
  \citenamefont {Lee}, \citenamefont {Kardar},\ and\ \citenamefont
  {Korolev}}]{Swartz2024}%
  \BibitemOpen
  \bibfield  {author} {\bibinfo {author} {\bibfnamefont {D.~W.}\ \bibnamefont
  {Swartz}}, \bibinfo {author} {\bibfnamefont {H.}~\bibnamefont {Lee}},
  \bibinfo {author} {\bibfnamefont {M.}~\bibnamefont {Kardar}},\ and\ \bibinfo
  {author} {\bibfnamefont {K.~S.}\ \bibnamefont {Korolev}},\ }\href
  {https://doi.org/10.48550/ARXIV.2405.19478} {\bibinfo {title} {New sector
  morphologies emerge from anisotropic colony growth}} (\bibinfo {year}
  {2024})\BibitemShut {NoStop}%
\bibitem [{\citenamefont {Hallatschek}\ and\ \citenamefont
  {Korolev}(2009)}]{Hallatschek2009}%
  \BibitemOpen
  \bibfield  {author} {\bibinfo {author} {\bibfnamefont {O.}~\bibnamefont
  {Hallatschek}}\ and\ \bibinfo {author} {\bibfnamefont {K.~S.}\ \bibnamefont
  {Korolev}},\ }\bibfield  {title} {\bibinfo {title} {Fisher waves in the
  strong noise limit},\ }\bibfield  {journal} {\bibinfo  {journal} {Physical
  Review Letters}\ }\textbf {\bibinfo {volume} {103}},\ \href
  {https://doi.org/10.1103/physrevlett.103.108103}
  {10.1103/physrevlett.103.108103} (\bibinfo {year} {2009})\BibitemShut
  {NoStop}%
\bibitem [{\citenamefont {Kardar}\ \emph {et~al.}(1986)\citenamefont {Kardar},
  \citenamefont {Parisi},\ and\ \citenamefont {Zhang}}]{Kardar1986}%
  \BibitemOpen
  \bibfield  {author} {\bibinfo {author} {\bibfnamefont {M.}~\bibnamefont
  {Kardar}}, \bibinfo {author} {\bibfnamefont {G.}~\bibnamefont {Parisi}},\
  and\ \bibinfo {author} {\bibfnamefont {Y.-C.}\ \bibnamefont {Zhang}},\
  }\bibfield  {title} {\bibinfo {title} {Dynamic scaling of growing
  interfaces},\ }\href {https://doi.org/10.1103/physrevlett.56.889} {\bibfield
  {journal} {\bibinfo  {journal} {Physical Review Letters}\ }\textbf {\bibinfo
  {volume} {56}},\ \bibinfo {pages} {889–892} (\bibinfo {year}
  {1986})}\BibitemShut {NoStop}%
\bibitem [{\citenamefont {Barabási}\ and\ \citenamefont
  {Stanley}(1995)}]{Barabsi1995}%
  \BibitemOpen
  \bibfield  {author} {\bibinfo {author} {\bibfnamefont {A.-L.}\ \bibnamefont
  {Barabási}}\ and\ \bibinfo {author} {\bibfnamefont {H.~E.}\ \bibnamefont
  {Stanley}},\ }\href {https://doi.org/10.1017/cbo9780511599798} {\emph
  {\bibinfo {title} {Fractal Concepts in Surface Growth}}}\ (\bibinfo
  {publisher} {Cambridge University Press},\ \bibinfo {year}
  {1995})\BibitemShut {NoStop}%
\bibitem [{\citenamefont {Moro}(2001)}]{Moro2001}%
  \BibitemOpen
  \bibfield  {author} {\bibinfo {author} {\bibfnamefont {E.}~\bibnamefont
  {Moro}},\ }\bibfield  {title} {\bibinfo {title} {Internal fluctuations
  effects on fisher waves},\ }\bibfield  {journal} {\bibinfo  {journal}
  {Physical Review Letters}\ }\textbf {\bibinfo {volume} {87}},\ \href
  {https://doi.org/10.1103/physrevlett.87.238303}
  {10.1103/physrevlett.87.238303} (\bibinfo {year} {2001})\BibitemShut
  {NoStop}%
\bibitem [{\citenamefont {Forster}\ \emph {et~al.}(1977)\citenamefont
  {Forster}, \citenamefont {Nelson},\ and\ \citenamefont
  {Stephen}}]{Forster1977}%
  \BibitemOpen
  \bibfield  {author} {\bibinfo {author} {\bibfnamefont {D.}~\bibnamefont
  {Forster}}, \bibinfo {author} {\bibfnamefont {D.~R.}\ \bibnamefont
  {Nelson}},\ and\ \bibinfo {author} {\bibfnamefont {M.~J.}\ \bibnamefont
  {Stephen}},\ }\bibfield  {title} {\bibinfo {title} {Large-distance and
  long-time properties of a randomly stirred fluid},\ }\href
  {https://doi.org/10.1103/physreva.16.732} {\bibfield  {journal} {\bibinfo
  {journal} {Physical Review A}\ }\textbf {\bibinfo {volume} {16}},\ \bibinfo
  {pages} {732–749} (\bibinfo {year} {1977})}\BibitemShut {NoStop}%
\bibitem [{\citenamefont {Takeuchi}(2017)}]{Takeuchi2017}%
  \BibitemOpen
  \bibfield  {author} {\bibinfo {author} {\bibfnamefont {K.~A.}\ \bibnamefont
  {Takeuchi}},\ }\bibfield  {title} {\bibinfo {title} {$1/f^\alpha$ power
  spectrum in the kardar–parisi–zhang universality class},\ }\href
  {https://doi.org/10.1088/1751-8121/aa7106} {\bibfield  {journal} {\bibinfo
  {journal} {Journal of Physics A: Mathematical and Theoretical}\ }\textbf
  {\bibinfo {volume} {50}},\ \bibinfo {pages} {264006} (\bibinfo {year}
  {2017})}\BibitemShut {NoStop}%
\bibitem [{\citenamefont {Nesic}\ \emph {et~al.}(2014)\citenamefont {Nesic},
  \citenamefont {Cuerno},\ and\ \citenamefont {Moro}}]{Nesic2014}%
  \BibitemOpen
  \bibfield  {author} {\bibinfo {author} {\bibfnamefont {S.}~\bibnamefont
  {Nesic}}, \bibinfo {author} {\bibfnamefont {R.}~\bibnamefont {Cuerno}},\ and\
  \bibinfo {author} {\bibfnamefont {E.}~\bibnamefont {Moro}},\ }\bibfield
  {title} {\bibinfo {title} {Macroscopic response to microscopic intrinsic
  noise in three-dimensional {F}isher fronts},\ }\bibfield  {journal} {\bibinfo
   {journal} {Physical Review Letters}\ }\textbf {\bibinfo {volume} {113}},\
  \href {https://doi.org/10.1103/physrevlett.113.180602}
  {10.1103/physrevlett.113.180602} (\bibinfo {year} {2014})\BibitemShut
  {NoStop}%
\bibitem [{\citenamefont {Easteal}(1981)}]{EASTEAL1981}%
  \BibitemOpen
  \bibfield  {author} {\bibinfo {author} {\bibfnamefont {S.}~\bibnamefont
  {Easteal}},\ }\bibfield  {title} {\bibinfo {title} {The history of
  introductions of bufo marinus (amphibia: Anura); a natural experiment in
  evolution},\ }\href {https://doi.org/10.1111/j.1095-8312.1981.tb01645.x}
  {\bibfield  {journal} {\bibinfo  {journal} {Biological Journal of the Linnean
  Society}\ }\textbf {\bibinfo {volume} {16}},\ \bibinfo {pages} {93–113}
  (\bibinfo {year} {1981})}\BibitemShut {NoStop}%
\bibitem [{\citenamefont {Shine}(2010)}]{Shine2010}%
  \BibitemOpen
  \bibfield  {author} {\bibinfo {author} {\bibfnamefont {R.}~\bibnamefont
  {Shine}},\ }\bibfield  {title} {\bibinfo {title} {The ecological impact of
  invasive cane toads (bufo marinus) in australia},\ }\href
  {https://doi.org/10.1086/655116} {\bibfield  {journal} {\bibinfo  {journal}
  {The Quarterly Review of Biology}\ }\textbf {\bibinfo {volume} {85}},\
  \bibinfo {pages} {253–291} (\bibinfo {year} {2010})}\BibitemShut {NoStop}%
\bibitem [{\citenamefont {Friedl}\ and\ \citenamefont
  {Alexander}(2011)}]{Friedl2011}%
  \BibitemOpen
  \bibfield  {author} {\bibinfo {author} {\bibfnamefont {P.}~\bibnamefont
  {Friedl}}\ and\ \bibinfo {author} {\bibfnamefont {S.}~\bibnamefont
  {Alexander}},\ }\bibfield  {title} {\bibinfo {title} {Cancer invasion and the
  microenvironment: Plasticity and reciprocity},\ }\href
  {https://doi.org/10.1016/j.cell.2011.11.016} {\bibfield  {journal} {\bibinfo
  {journal} {Cell}\ }\textbf {\bibinfo {volume} {147}},\ \bibinfo {pages}
  {992–1009} (\bibinfo {year} {2011})}\BibitemShut {NoStop}%
\bibitem [{\citenamefont {Wirtz}\ \emph {et~al.}(2011)\citenamefont {Wirtz},
  \citenamefont {Konstantopoulos},\ and\ \citenamefont {Searson}}]{Wirtz2011}%
  \BibitemOpen
  \bibfield  {author} {\bibinfo {author} {\bibfnamefont {D.}~\bibnamefont
  {Wirtz}}, \bibinfo {author} {\bibfnamefont {K.}~\bibnamefont
  {Konstantopoulos}},\ and\ \bibinfo {author} {\bibfnamefont {P.~C.}\
  \bibnamefont {Searson}},\ }\bibfield  {title} {\bibinfo {title} {The physics
  of cancer: the role of physical interactions and mechanical forces in
  metastasis},\ }\href {https://doi.org/10.1038/nrc3080} {\bibfield  {journal}
  {\bibinfo  {journal} {Nature Reviews Cancer}\ }\textbf {\bibinfo {volume}
  {11}},\ \bibinfo {pages} {512–522} (\bibinfo {year} {2011})}\BibitemShut
  {NoStop}%
\bibitem [{\citenamefont {Pechenik}\ and\ \citenamefont
  {Levine}(1999)}]{Pechenik1999}%
  \BibitemOpen
  \bibfield  {author} {\bibinfo {author} {\bibfnamefont {L.}~\bibnamefont
  {Pechenik}}\ and\ \bibinfo {author} {\bibfnamefont {H.}~\bibnamefont
  {Levine}},\ }\bibfield  {title} {\bibinfo {title} {Interfacial velocity
  corrections due to multiplicative noise},\ }\href
  {https://doi.org/10.1103/physreve.59.3893} {\bibfield  {journal} {\bibinfo
  {journal} {Physical Review E}\ }\textbf {\bibinfo {volume} {59}},\ \bibinfo
  {pages} {3893–3900} (\bibinfo {year} {1999})}\BibitemShut {NoStop}%
\bibitem [{\citenamefont {Dornic}\ \emph {et~al.}(2005)\citenamefont {Dornic},
  \citenamefont {Chaté},\ and\ \citenamefont {Muñoz}}]{Dornic2005}%
  \BibitemOpen
  \bibfield  {author} {\bibinfo {author} {\bibfnamefont {I.}~\bibnamefont
  {Dornic}}, \bibinfo {author} {\bibfnamefont {H.}~\bibnamefont {Chaté}},\
  and\ \bibinfo {author} {\bibfnamefont {M.~A.}\ \bibnamefont {Muñoz}},\
  }\bibfield  {title} {\bibinfo {title} {Integration of langevin equations with
  multiplicative noise and the viability of field theories for absorbing phase
  transitions},\ }\bibfield  {journal} {\bibinfo  {journal} {Physical Review
  Letters}\ }\textbf {\bibinfo {volume} {94}},\ \href
  {https://doi.org/10.1103/physrevlett.94.100601}
  {10.1103/physrevlett.94.100601} (\bibinfo {year} {2005})\BibitemShut
  {NoStop}%
\end{thebibliography}
\end{document}